# Large, high quality single-crystals of the new Topological Kondo Insulator, SmB$_6$.

M. Ciomaga Hatnean, M. R. Lees, D. M$^c$K. Paul and G. Balakrishnan*

Department of Physics, University of Warwick, Coventry, CV4 7AL, UK

## ABSTRACT

SmB$_6$ has recently been predicted to be a Topological Kondo Insulator, the first strongly correlated heavy fermion material to exhibit topological surface states. High quality crystals are necessary to investigate the topological properties of this material. Single crystal growth of the rare earth hexaboride, SmB$_6$, has been carried out by the floating zone technique using a high power xenon arc lamp image furnace. Large, high quality single-crystals are obtained by this technique. The crystals produced by the floating zone technique are free of contamination from flux materials and have been characterised by resistivity and magnetisation measurements. These crystals are ideally suited for the investigation of both the surface and bulk properties of SmB$_6$.

* Corresponding author, E-mail: G. Balakrishnan@warwick.ac.uk

The rare earth hexaborides ($RB_6$) have been studied for their interesting properties for many decades. Magnetic rare earth hexaborides such as $CeB_6$ have been investigated in great detail [1, 2], whilst $LaB_6$ is of interest for its extraordinary electron emitting properties and use in Electron Microscopes [3, 4]. Of the other rare earth hexaborides, $SmB_6$ is known to be a Kondo insulator and has been a material that has intrigued physicists for many years [5, 6]. Kondo insulators have been known for a long time, and are heavy electron (fermion) materials where the local magnetic moments are screened by the itinerant electrons. The hybridisation of the conduction electrons with the magnetic moments below a certain temperature leads to the formation of an energy gap in the density of states [7, 8].

Following recent theoretical work by Dzero *et al*. it has been suggested that $SmB_6$ is a Topological Kondo Insulator exhibiting topological surface properties [9, 10, 11, 12, 13]. As a result, there has been renewed interest in this previously well studied Kondo insulator. $SmB_6$ exhibits an insulator to metal transition at around 50 K and this coincides with the opening of the gap due to the Kondo behaviour [14, 15, 16]. At low temperatures (<5 K) the resistivity shows a saturation behaviour unlike in most insulators, where the resistivity continues to rise. At ambient pressures, $SmB_6$ is a mixed valent material exhibiting antiferromagnetic correlations, but no magnetic ordering. When subjected to pressures, $SmB_6$ is known to go into a metallic, magnetically ordered state [17, 18].

Since the proposal that $SmB_6$ could exhibit topological behaviour, there has been intense activity amongst experimentalists looking for evidence of topological surface states through angle resolved photoemission (ARPES) experiments, and other related surface techniques [18, 19, 20, 21, 22]. An estimation of the Kondo insulating gap has also been attempted by tunnelling techniques [23].

In order to observe the topological surface states and perform detailed experiments on these materials, high quality single crystals are a prerequisite. For the crystal growth of hexaborides, two different methods are used: (i) crystal growth by the flux technique, using Al flux, as has been successfully carried out for the whole family of rare earth hexaborides [24, 25] and (ii) crystal growth by the floating zone technique using either R.F. heating or lamp heating. The rare earth hexaboride materials have very high melting temperatures (~ 2500 °C and above) and therefore melting these materials requires the use of Xenon arc lamps, if using an optical furnace, to reach these high temperatures. The flux-grown crystals are usually very small in size (~1 mm edge) and most of the recent experimental work performed and published on $SmB_6$ has been on crystals produced by this technique.

For many studies however, large, clean, defect free crystals are preferred and the floating zone technique is the ideal route to prepare such samples. Crystals of some hexaborides including $SmB_6$ have been grown from a melt using R.F. heating [26]. This paper describes the crystal growth and characterisation of large, high quality single-crystals of $SmB_6$ by the floating zone technique using Xenon arc lamps. The crystals obtained are free from any contamination from fluxes or crucibles and are ideal for the crucial experiments necessary to establish the true behaviour of both the surfaces and the bulk of $SmB_6$.

**Results**

$SmB_6$ is known to melt congruently, similar to $LaB_6$, $CeB_6$ and $NdB_6$ [26]. Figure 1(a) shows a photograph of a portion of a crystal boule of $SmB_6$ grown by the floating zone method. The obtained crystals were 7 to 8 mm in diameter and about 20 to 25 mm long. The maximum power of the lamps needed to maintain a stable molten zone throughout the growth was around 70%. Excessive evaporation of the $SmB_6$ prevented us from maintaining a stable molten zone for longer than the 25 to 30 mm of growth. This is unlike the growth conditions encountered by us for the other congruently melting hexaborides, where longer growths were possible [27]. Growths were also attempted using high gas pressures (up to 0.7 MPa argon gas pressure) to suppress the evaporation, but no improvement in the growth conditions was observed. The lamp power used for the growth of the $SmB_6$ crystals is below the ~85% required for the growth of $LaB_6$ using the same furnace and we therefore believe that growth of larger diameter crystals is possible by this technique. The crystal boules developed well defined facets, within the first few millimeters of the growth and the boules obtained had a dark black colour. The X-ray Laue patterns of the as-grown crystal boules indicate that the crystals were of good quality (see Figure 1). The Laue patterns were taken along the length of the boule, from the faceted sides (shown in Fig.1(a)) . Identical patterns were obtained along the whole length of the faceted faces. Laue patterns taken from the cross section of the boule at the tip, show that the crystal growth direction was only a few degrees away from the [110] direction (see Figure 1(b)). Consistent Laue patterns were obtained at several positions on the cross section, two of which are shown in Fig. 1(a).

Due to the extreme hardness of the borides in general, the crystals needed to be cut using the spark erosion technique. The cut surfaces, when cleaned with very dilute acid to remove the copper deposits from the spark erosion process, reveal very shiny faces with a deep purple hue. Composition analysis by EDAX of the crystals produced show that the composition sampled over a large area of the crystals averages to 1:6 for Sm:B, despite the observed evaporation during the crystal growth.

The powder X-ray diffraction pattern obtained on a crushed crystal sample cut from the as-grown $SmB_6$ crystal boule is shown in Figure 2. The pattern obtained showed no additional peaks due to any secondary phases, confirming that the crystals were single phase to within the limit of the X-ray diffraction technique. The observed pattern has been analysed using the FullProf software suite [28] and the results of the Rietveld refinements show that all the observed peaks could be indexed to the cubic *Pm3m* space group. The obtained value of 4.1353(1) Å for the lattice parameter is in agreement with the previously published value for $SmB_6$ [29].

Figure 3 shows the resistivity measurements made on a bar shaped sample cut from the $SmB_6$ crystal. The temperature dependence and magnitude of the resistivity, as well the temperature at which there is an upturn in the resistivity indicating the onset of the transition to the insulating state, agree with previously reported data collected on the flux grown crystals [30, 31]. The flattening of the resistance observed at low temperatures (shown in the inset) due to the saturation in the resistivity is attributed to a conducting channel, which is explained as being due to the existence of topologically protected surface states. This saturation is a feature observed in the $SmB_6$ crystals produced both by the flux technique as well as here by the floating zone technique.

The temperature dependence of the magnetisation in different applied magnetic fields for a piece of the single crystal is shown in Fig. 4. The increase in the resistivity just below 60 K coincides with a very sharp rise in the susceptibility seen in the measurement made in low field (500 Oe). This is followed by a broad maximum at lower temperature and an upturn at base temperature. Data collected in higher fields (5 and 20 kOe) exhibit only the broad hump, which shifts to higher temperature with increasing field, and the low temperature upturn. The temperature dependence of the susceptibility of our crystals is consistent with what is expected for this Kondo insulator. The features seen in our low field magnetic susceptibility measurements are much sharper than those reported previously [32, 33, 34].

**Discussion**

The congruent melting property of $SmB_6$ facilitates the growth of single crystals from the melt. The sizes of the crystals grown by the optical floating zone technique are much larger than those that can be obtained by the flux technique. The crystalline quality of the boules as examined by Laue X-ray diffraction patterns indicate that the faceted faces of the crystal boules exhibit identical patterns, on the two faceted faces, at 180 degrees with respect to each other.

As a comparison, we have also produced $SmB_6$ crystals via the standard aluminium flux technique [25]. The crystals have well defined shiny faces and the maximum sizes produced are roughly 0.5 to 1~ mm edge. Whilst these may be easier to produce with flat surfaces required for some studies, they often have contamination from the flux materials used. The crystals produced by the optical floating zone technique, in addition to being much bigger, are free of any inclusions and of extremely good quality. Another advantage of these large crystals is that specimens aligned along specific crystallographic directions can be cut from the boule for experiments.

To summarise, we have successfully produced large single crystals of the new Topological Kondo Insulator, $SmB_6$, by the floating zone technique using a Xenon arc lamp furnace. Examination of the crystals with X-ray Laue diffraction indicates that the quality of the crystals is good. The crystals obtained are free of any contamination when examined by powder X-ray diffraction on the crushed crystals. They exhibit all the hallmarks of a Kondo Insulator in both resistivity and magnetisation measurements. The crystals produced are ideal for the investigation of both surface as well as bulk properties, to understand the existence of topological surface states in this interesting Kondo insulator. Measurements of the low temperature properties on crystal specimens oriented along specific crystallographic axes can be performed using these large crystals, and these experiments are currently under progress. Crystals can also be obtained by the same route using isotopically enriched $^{11}B$ as shown previously by us for other members of the hexaboride family [27]. For neutron scattering investigations, however, in addition to using the $^{11}B$ isotope, the highly absorbing Sm would also need to be replaced with the less absorbing $^{154}Sm$ isotope.

**Methods**

Commercial samarium hexaboride powder (99.9% Cerac, USA) was used as starting material for making the feed and seed rods. The powder was mixed with a small amount of PVA or PVB binder, ground well and pressed into rods of approximately 6 mm diameter and 60 to 70 mm length. The resulting rods were sintered in a furnace in a flow of argon gas at 1550 °C for 12 h. Before starting the sintering process, the furnace was evacuated to give a vacuum of ~$10^{-5}$ mbar (~$10^{-3}$ Pa). The sintered rods were used for the crystal growth. The crystal growth was carried out by the floating zone method, using a Xenon arc lamp furnace (CSI FZ-T-12000-X-VI-VP, Crystal Systems Incorporated, Japan). This furnace has four arc lamps with a total output power of 12 kW and is capable of reaching temperatures close to ~2800°C. The crystal growth was carried out in about 3 bars argon gas pressure and a flow of argon gas of up to 10 l/min, using a maximum growth rate of ~18 mm/h. Both the feed and the seed rods were counter-rotated at 30 rpm to ensure efficient mixing and homogeneity. Polycrystalline rods were used as seeds for the first growth and the crystals obtained were used as seeds for the subsequent growths.

The crystal boules produced were first examined using X-ray Laue diffraction to check the quality of the grown $SmB_6$ crystals. Spark erosion was used to cut slices from the grown boule for subsequent measurements. Powdered samples of single crystal boules were analysed using X-ray diffraction (XRD) performed at room temperature in a Panalytical X-ray diffractometer using Cu K$\alpha_1$ radiation ($\lambda$ = 1.5406 Å). The X-ray data were analysed with the FullProf software suite [28]. Composition analysis on the crystals grown was carried out by EDAX on a scanning electron microscope.

Temperature dependent magnetisation measurements were made in the temperature range 1.8 to 300 K, and in applied magnetic fields of 500 Oe, 5 kOe and 20 kOe, using a Quantum Design MPMS-5S SQUID Magnetometer. DC resistivity measurements were carried out on rectangular bar shaped samples by the standard four probe technique using a Quantum Design Physical Property Measurement System (PPMS) in the temperature range 1.8 to 300 K.

**References:**

1. Effantin, J. M. *et al*. Magnetic Phase diagram of CeB$_6$. *J. Magn. Magn. Mater.* **47-48**, 145-148 (1985).

2. Regnault, L. P. *et al*. Inelastic neutron scattering study of rare earth hexaboride CeB$_6$. *J. Magn. Magn. Mater*. **76-77**, 413-414 (1988).

3. Tanaka, T., Bannai, E., Kawai, S. & Yamane, T. Growth of high purity LaB$_6$ crystals by the multi-float zone passage. *J. Crys. Growth* **30** 193-197 (1975).

4. Otani, S., Hiraoka, H., Ide, M. & Ishizawa, Y. Thermionic emission properties of rare earth added $LaB_6$ crystal cathodes, *J. Alloy. Compd.* **189**, L1-L3 (1992).

5. Aeppli, G. & Fisk, Z. Kondo Insulators. *Comments Condens. Matter Phys*. **16**, 155-165 (1992).

6. Riseborough, P. Heavy Fermion semiconductors. *Adv. Phys*. **49**, 257-320 (2000).

7. For a recent review, see Coleman, P. Heavy Fermions: Electrons at the edge of magnetism. *Handbook of Magnetism and Advanced Magnetic Materials* (Wiley, New York, 2007). Vol. **1** 95-148.

8. Wachter, P. *Handbook on the Physics and Chemistry of Rare Earths* Vol. **19** (*Eds. K. A. Gsneidner, Jr and L. Eyring),* (North Holland, Amsterdam, 1994).

9. Dzero, M., Sun, K., Galitski, V. & Coleman, P. Topological Kondo Insulators. *Phys. Rev. Lett.* **104** 106408 (2010).

10. Dzero, M., Sun, K., Coleman, P. & Galitski, V. A theory of Topological Insulators. *Phys. Rev.* B **85**, 045130 (2012).

11. Fu, L., Kane, C. L. & Mele, E. J. Topological Insulators in three dimensions. *Phys. Rev. Lett.* **98**, 106803 (2007).

12. Moore, J. E. & Balents, L. Topological invariants of time reversal invariant band structures. *Phys. Rev.* B **75**, 121306 (R) (2007).

13. Roy, R. Topological phases and the quantum spin Hall effect in three dimensions. *Phys. Rev.* B **79**, 195322 (2009).

14. Menth, A., Buehler, E. & Geballe, T. Magnetic and semiconducting properties of $SmB_6$, *Phys. Rev. Lett*. **22**, 295-297 (1969).

15. Allen, J., Batlogg, B. & Wachter, P. Large low temperature Hall effect and resistivity in mixed valent $SmB_6$. *Phys. Rev*. B **20**, 4807-4813 (1979).

16. Cooley, J., Aronson, M., Fisk, Z. & Canfield P. $SmB_6$: Kondo insulator or exotic metal? *Phys. Rev. Lett*. **74**, 1629 (1995).

17. Barla, A. *et al*. High pressure ground state of SmB$_6$: Electronic conduction and long range magnetic order. *Phys. Rev. Lett*. **94**, 166401 (2005).

18. Miyazaki, H., Hajiri, T., Ito, T., Kunii, S. & Kimura, S. Momentum dependent hybridization gap and dispersive in-gap state of the Kondo semiconductor SmB$_6$. *Phys. Rev.* B **86**, 075105 (2012).

19. Xu, N. et al. Surface and bulk electronic structure of the strongly correlated electron system SmB$_6$ and implications for a topological Kondo insulator. arXiv:1306.3678.

20. Jiang, J. *et al*. Observation of in-gap surface states in the Kondo insulator SmB$_6$ by photoemission. arXiv:1306.5664.

21. Neupane, M. *et al*. Surface electronic structure of a topological Kondo insulator candidate SmB$_6$: insights from high-resolution ARPES. arXiv:1306.4634.

22. Frantzeskakis, E. *et al*. Kondo hybridisation and the origin of metallic states at the (001) surface of SmB$_6$. arXiv:1308.0151.

23. Yee, M. M. *et al*. Imaging the Kondo Insulating gap in SmB$_6$. arXiv:1308:1085.

24. Canfield, P. C. & Fisk, Z. Growth of single crystals from metallic fluxes. *Phil. Mag. Part B* **65**, 1117-1123 (1992).

25. Fisk, Z. *et al*. Magnetic transport and thermal properties of ferromagnetic EuB$_6$. *J. Appl. Phys*. **50**, 1911 (1979).

26. Otani, S., Nakagawa, H., Nishi, Y. & Kieda, N. Floating Zone Growth and high temperature hardness of rare earth hexaboride crystals: LaB$_6$, CeB$_6$, PrB$_6$, NdB$_6$ and SmB$_6$. *J. Solid State Chem*. **154**, 238-241 (2000),

27. Balakrishnan, G., Lees, M. R. Paul, D. McK. Growth of large single crystals of rare earth hexaborides. *J. Cryst. Growth* **256**, 206-209 (2003).

28. Rodriguez-Carvajal J. Recent advances in the magnetic structure determination by neutron powder diffration. *Phys. B: Condens. Matter* **192**, 55-69 (1993).

29. Tarascon, J. M., Isikawa, Y., Chevalier, B., Etourneau, J. & Hagenmuller, P. Valence transition of samarium in hexaboride solid solutions Sm$_{1-x}$M$_x$B$_x$ (M = Yb$^{2+}$, Sr$^{2+}$, La$^{3+}$, Y$^{3+}$, Th$^{4+}$). *J. Physique*

**41**, 1135-40 (1980).

30. Kim, D. J., Grant, T. & Fisk, Z. Limit cycle and anomalous capacitance in the Kondo Insulator *Phys. Rev. Lett.* **109**, 096601 (2012).

31. Flachbart, K. *et al*. Energy gap of intermediate valent $SmB_6$ studied by point-contact spectroscopy. *Phys. Rev. B* **64**, 085104 (2001).

32. Yeo, S., Song, K., Hur, N., Fisk, Z. & Schlottmann, P. Effects of Eu doping on $SmB_6$ single crystals, *Phys. Rev.* B **85**, 115125 (2012).

33. Gabani, S. *et al*. J. Magnetic properties of $SmB_6$ and $Sm_{1-x}La_xB_6$ solid solutions. *Czech. J. Phys.* **52**, A225-228 (2002).

34. Glushkov, V. V. *et al*. Spin gap formation in $SmB_6$. *Physica B* **378-380**, 614-615 (2006).

**Acknowledgments**:

This work was supported by EPSRC, UK through Grant EP/I007210/1. We thank T. E. Orton for valuable technical support and S. J. York for the composition analysis. GB wishes to thank Prof. Takaho Tanaka (NIMS, Japan) for helpful discussions. Some of the equipment used for the measurements was obtained through the Science City Advanced Materials: Creating and Charaterising Next Generation Advanced Materials Project, with support from Advantage West Midlands (AWM) and part funded by the European Regional Development Fund (ERDF).

**Author Contributions**:

GB conceived of the project. MCH and GB performed the crystal growth. MCH performed the characterisation measurements with MRL and analysed the data. GB drafted the paper with significant contributions from MCH and MRL. MCH, MRL and DMP reviewed the manuscript. All others who contributed to this project are recognised in the Acknowledgements.

**Additional Information**:

Competing financial interests: The authors declare no competing financial interest.

**Figure captions**

**Fig. 1 (a) Photograph of a portion of an as-grown boule of $SmB_6$ prepared by the floating zone technique.** Also shown below the image of the crystal are the Laue patterns of one of the facets, taken along the crystal length at 4-6 mm intervals. The corresponding Laue patterns taken on the facet at 180 degrees are mirror images of these patterns. Laue patterns taken of the cross section of the boule at the tip, are consistent with one another at several points covering the entire surface. Two patterns corresponding to a scan of the top and bottom of the tip, are shown (on the left).
(b) X-ray Laue back reflection photograph of a crystal of $SmB_6$, along the [110] direction.

**Fig. 2. Powder X-ray diffraction pattern of a crushed sample of the as-grown $SmB_6$ crystal.** The data obtained (red), the fit to the data using Rietveld refinement (black line), as well as the difference curve (blue line) are shown.

**Fig. 3. Temperature dependence of the dc resistivity for a single crystal of $SmB_6$.** The resistivity increases rapidly below 60 K before flattening at temperatures below 3.5 K (see inset).

**Fig. 4. Magnetic susceptibility (*M/H)* as a function of temperature for a single crystal of $SmB_6$**. The data for an applied field of 500 Oe were collected on warming after cooling in zero-field (ZFCW, open symbols) and on cooling in the same field (FCC, closed symbols). No significant

hysteresis is seen between these two data sets. Data were also collect in FCC mode in applied fields of 5 and 20 kOe.

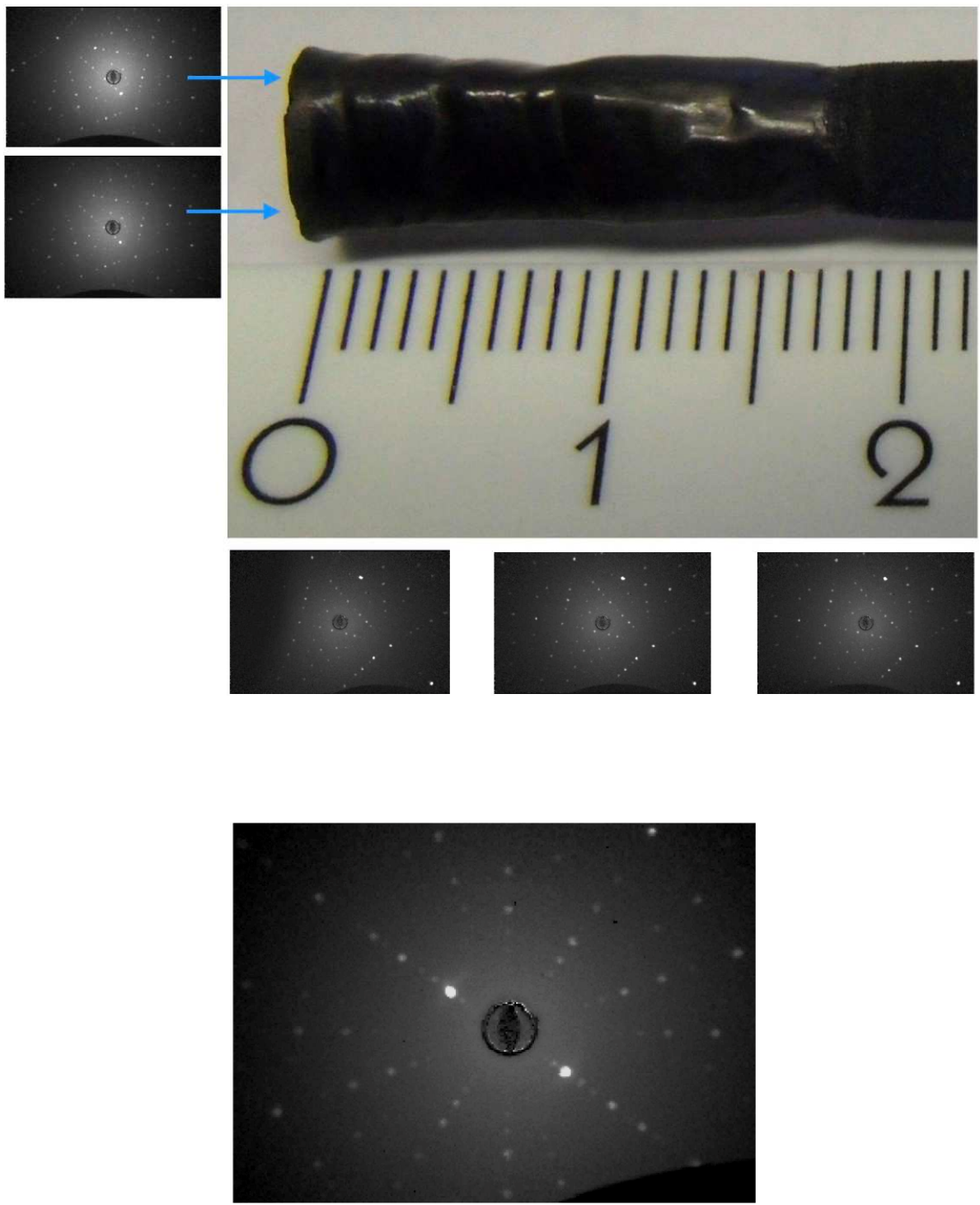

Fig. 1.

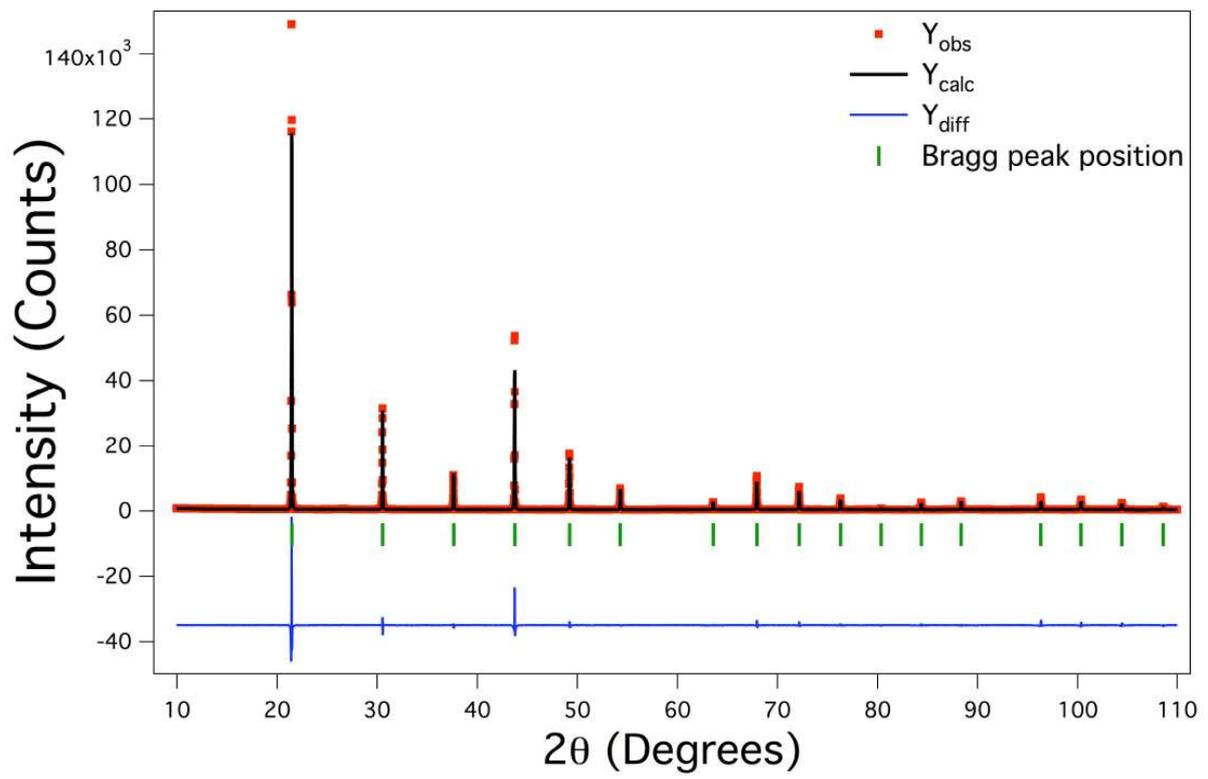

Fig. 2

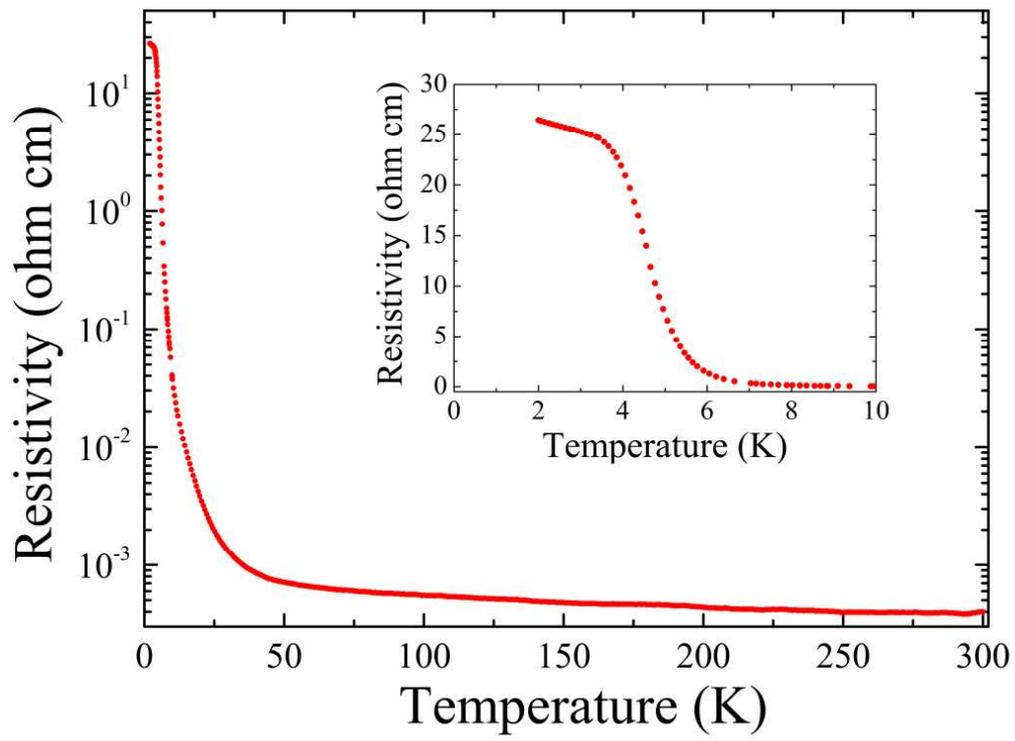

Fig. 3

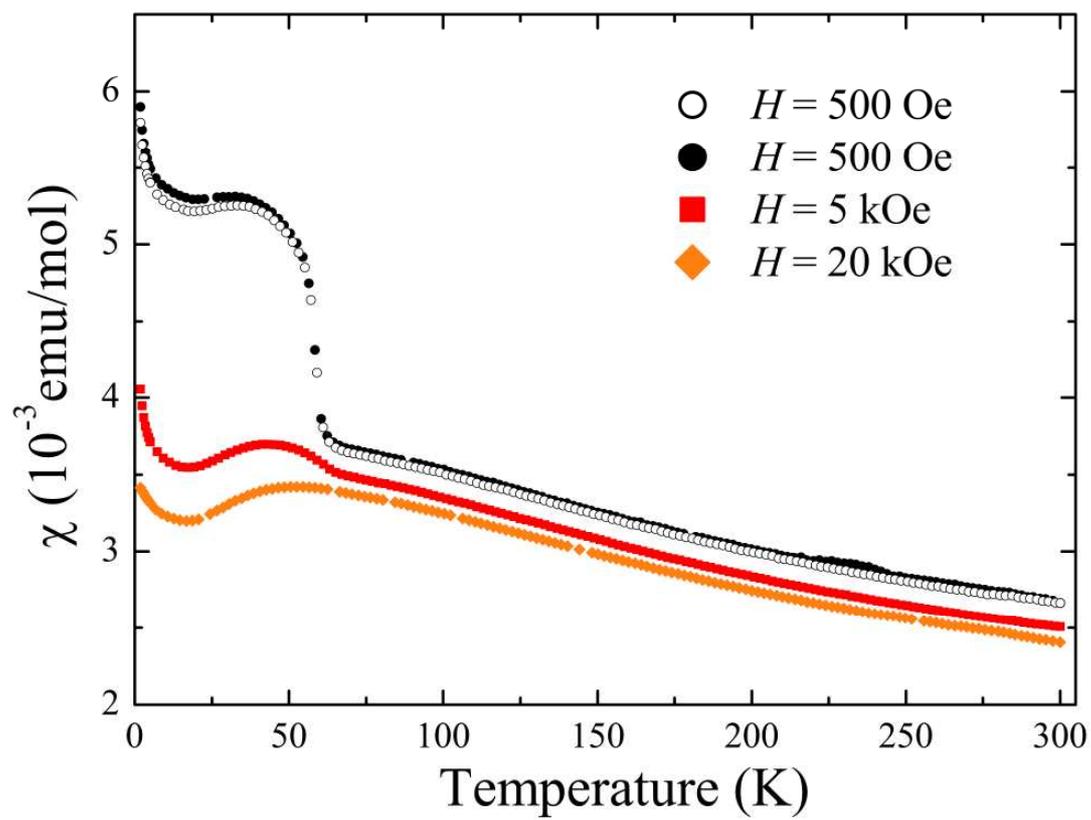

Fig. 4